\documentclass{aa}  
\usepackage{graphicx}
\usepackage{epstopdf}
\usepackage{comment}
\usepackage[varg]{txfonts}
\begin{document}
   \title{Refined physical properties and g',r',i',z',J,H,K transmission
   spectrum of WASP-23b from the ground\thanks{Based on observations collected with the Gamma Ray Burst
Optical and Near-Infrared Detector (GROND) at the MPG/ESO-2.2m
telescope at La Silla Observatory, Chile. Programme 088.A-9006}}
%   \subtitle{I. A Ground-based Broad-band Transmission Spectrum of WASP-23b}

   \author{N. Nikolov
          \inst{1,}\inst{2}
          \and G. Chen\inst{1,3}           
          \and J. J. Fortney\inst{4} 
          \and L. Mancini\inst{1}
          \and J. Southworth\inst{5} 
          \and R. van Boekel\inst{1} 
          \and Th. Henning\inst{1}
          }
   \institute{Max Planck Institute for Astronomy, 
   K\"{o}nigstuhl 17, 69117 -- Heidelberg, Germany
         \and
             Astrophysics Group, School of Physics, University of Exeter, Stocker Road, EX4 4QL, Exeter, UK\\
              \email{nikolay@astro.ex.ac.uk}
         \and
            Purple Mountain Observatory \& Key Laboratory for Radio Astronomy, 2 West Beijing Road, Nanjing 210008, China
         \and
            Department of Astronomy \& Astrophysics, University of California, Santa Cruz, CA 95064, USA
         \and
            Astrophysics Group, Keele University, Newcastle-under-Lyme, ST5 5BG, UK
             }

   \date{Received: January 11, 2013; Accepted: March 14, 2013.}%Received November 15, 2012; accepted....

% \abstract{}{}{}{}{} 
% 5 {} token are mandatory
 
%{ \bf{}}
 
%==============================================================================================================
   \abstract
   {Multi-band observations of planetary transits using the telescope defocus technique may yield high-quality light curves suitable for refining the physical properties of exoplanets even with small or medium size telescopes. Such observations can be used to construct a broad-band transmission spectrum of transiting planets and search for the presence of strong absorbers.}
   {We have thoroughly characterised the orbital ephemeris and physical properties of the transiting planet and host star in the {\rm{WASP$-$23b}} system, constructed a broad-band transmission spectrum of \object{\object{{\rm{WASP$-$23\,b}}}} and performed a comparative analysis with theoretical models of hot Jupiters.}
   {We observed a complete transit of \object{\object{{\rm{WASP$-$23\,b}}}} in seven passbands simultaneously, using the GROND instrument on the MPG/ESO 2.2\,m telescope at La Silla Observatory and telescope defocussing. The optical data were taken in the {\it{Sloan}} g$^{\prime}$, r$^{\prime}$, i$^{\prime}$ and z$^{\prime}$ passbands. The resulting light curves are of high quality, with a root-mean-square scatter of the residual as low as 330 parts per million (ppm) in the z$^{\prime}$-band, with a cadence of 90\,s. Near-infrared data were obtained in the JHK passbands. We performed a MCMC analysis of our photometry plus existing radial velocity data to refine measurements of the ephemeris and physical properties of the {\rm{WASP$-$23}} system. We constructed a broad-band transmission spectrum of \object{\object{{\rm{WASP$-$23\,b}}}} and compared it with a theoretical transmission spectrum of a Hot Jupiter.}
   {We measured the central transit time with a precision $\sim$8\,s. From this and earlier observations we obtain an orbital period of $P = 2.9444300 \pm 0.0000011$\,d. Our analysis also yielded a larger radius and mass for the planet ($R_{p}=1.067^{+0.045}_{-0.038}$ ${\rm{R}}_{Jup}$ and $M_{p}=0.917^{+0.040}_{-0.039}$ ${\rm{M}}_{Jup}$) compared to previous estimates ($R_{p}=0.962^{+0.047}_{-0.056}$ ${\rm{R}}_{Jup}$ and $M_{p}=0.884^{+0.088}_{-0.094}$ ${\rm{M}}_{Jup}$). The derived transmission spectrum is marginally flat, which is not surprising given the limited precision of the measurements for the planetary radius and the poor spectral resolution of the data. }
   {}%Our results imply that the radius of WASP-23b is not inflated
%==============================================================================================================


   \maketitle

%==============================================================================================================
\section{Introduction}
%==============================================================================================================

The discovery and the detailed analysis of planets orbiting other stars can yield much information about their physical properties and evolutionary pathways. Since the first detection of a Jupiter-like planet hosted by a main sequence star \citep{mayor_queloz1995}, several hundred exoplanets have been discovered via different methods. These comprise radial velocity (RV), astrometry, transits, microlensing, direct imaging and timing (see \cite{schneider_et_al2011}). Among the currently known exoplanets, those that transit their host stars command great interest because so many of their physical properties can be determined \citep{seager2003, torres_et_al2010, sozzetti_et_al2007}. The properties of particular value include their radii and masses, and hence their surface gravities and mean densities. Observations of planetary transits currently provide the best route to constructing the mass-radius diagram of exoplanets, which brings critical information on the structure, formation and evolution of these worlds \citep{baraffe_2010, fortney_et_al_2007a, burrows_et_al2007}. Moreover, accurate measurements of the times of transit midpoints can allow the detection of additional bodies in known exoplanetary systems \citep{holman_murray2005} or alternatively may serve as a tool to verify the planetary nature and study the dynamical effect in multi-planet systems \citep{lissauer_et_al2011}. 

Notably, exoplanetary transits provide an opportunity to study the atmospheres of these objects. During transits part of the star light filters through the planetary atmosphere, and is imprinted with characteristic signatures of atomic (such as Na and K) and molecular (e.g.\ ${\rm{H_{2}O}}$, CO and ${\rm{CH_{4}}}$) absorption \citep{seager_sasselov2000,hub2001}.

The amount of irradiation incident upon giant extrasolar planets is an important factor in determining the properties of their atmospheres.  One theory of giant exoplanets at high incident stellar fluxes predicts the existence of the pM and pL classes of hot Jupiters, depending on the presence of strong absorbers, such as gaseous TiO and VO in their atmospheres \citep{fortney2008, fortney2010}. In particular, the increased opacity due to TiO and VO in the pM class should result in about a $3\%$ variation of the observed radius of a planet with a surface gravity of 15\,m\,s$^{-2}$, over the wavelength interval 350--700\,nm. For the purposes of transmission spectroscopy it is practical to convert these planet variations to the change of a directly observed quantity such as the transit depth $\delta = (R_{{\rm{p}}}/R_{{\rm{\ast}}})^2$. Assuming a radius variation with an amplitude $n=0.03$ (3\%), the transit depth variation is defined as
\begin{equation}
\Delta \delta = \frac{\pi(R_{{\rm{p}}} + nR_{{\rm{p}}})^2}{ \pi R^2_{\ast}} -\frac{ \pi R^2_{{\rm{p}}}}{  \pi R^2_{\ast}} \approx 2 n \delta
\end{equation}
A transiting Jupiter orbiting a Sun-like star causes $\delta = 0.01$ which would translate into a variation of $\Delta \delta = 6\times10^{-4}$ (or $\Delta (R_{{\rm{p}}}/R_{\ast}) = \sqrt{2 n \delta} \approx 0.024$). Such precisions may be achieved by modelling light curves obtained using ground-based 2-4\,m telescopes. The best results can be obtained by defocusing the telescope, which averages out sensitivity variations between the pixels on the detector and improves the observing efficiency by enabling a larger fraction of open-shutter time. However, the bandwidths of the filters used (which give the spectral resolution) is the other critical factor affecting the chances of detecting the signal in Equation 1, similar to the detection of spectral features in spectrophotometry.

Several successful searches for the theoretically predicted strong absorbers in the optical and near-infrared (NIR) have been conducted first from space using the {\it{Hubble}} and {\it{Spitzer}} space telescopes, resulting in the detection of Na, CO, H$_2$O, CH$_4$ and Rayleigh scattering in the blue \citep{charbonneau2002,  pont2008, sing2011b}. Hampered by the atmosphere of the Earth (e.g.\ telluric contamination) and instrumental systematics, the ground-based detection of spectral features in exoplanet atmospheres took longer to yield results \citep{redfield2008, snellen2008, sing2011a, sing2012}.

%%%%%%%%%%%%%%%%%%%%%%%%%%%%%%%%%%%%%%%%%%
%                                                FIGURE 1                                                       %
%%%%%%%%%%%%%%%%%%%%%%%%%%%%%%%%%%%%%%%%%%
   \begin{figure}[!t]
   \centering
   \includegraphics[width=\hsize]{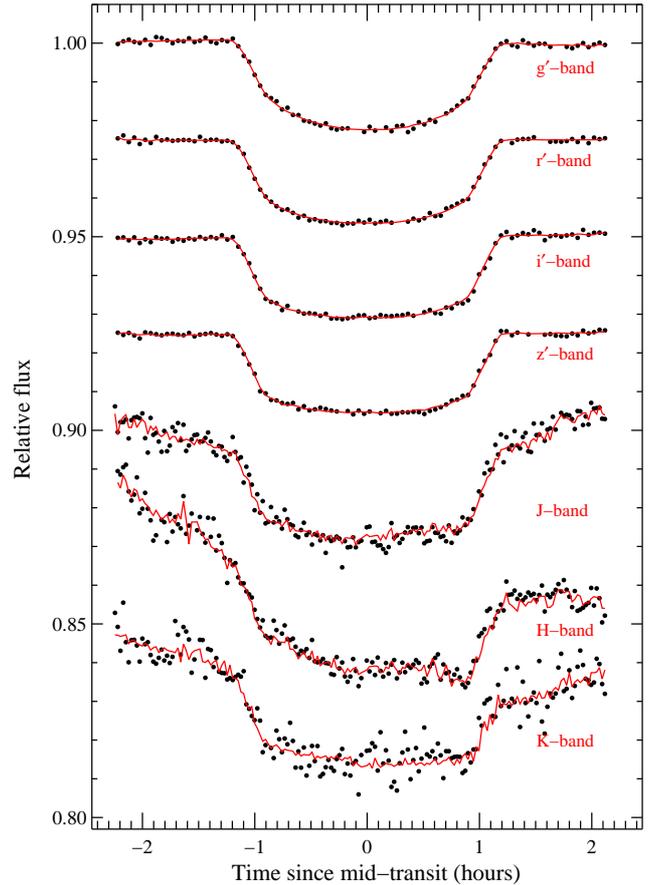}
      \caption{Simultaneous optical to near-infrared ({\it{Sloan}} g$'$,r$'$,i$'$,$'z$ and {\it{Johnson}} J, H, K) raw transit light curves of {{WASP$-$23}} (dots), obtained with GROND on UT February 25, 2012 an the corresponding correction functions (red lines).}
         \label{FigVibStab}
   \end{figure}
%%%%%%%%%%%%%%%%%%%%%%%%%%%%%%%%%%%%%%%%%%%
 
Although transit photometry or spectrophotometry is often of high quality, the method pushes the limits of current instrumentation and has yielded controversial results \citep{swain2008, beaulieu2011}. The success of the method may also be hampered by the activity of the host star, in particular spot crossing events which can significantly increase the complexity of light curve analysis and lower the precision of the planetary radius measurement \citep{knutson2011}.
 
The transiting hot Jupiter \object{{\rm{WASP$-$23\,b}}} orbits a moderately bright ($\rm{V} = 12.68$) K1\,V ($\rm{T}_{\rm{eff}} = 5150 \pm 100\,\rm{K}$) 
star and was found by the SuperWASP survey \citep{t11, pollacco2006}. The planet (radius $0.96 \pm 0.05\,R_{\rm{J}}$, mass $0.88 \pm 0.10\,M_{\rm{J}}$)
has a 2.94\,d circular orbit ($e < 0.062$ at the $3\sigma$ level).

In this paper we present ground-based simultaneous optical and NIR photometry of a transit of {{WASP$-$23\,b}}. We refine its measured properties and orbital ephemeris. We then construct the first transmission spectrum of {{WASP$-$23\,b}} and investigate the variation in radius of the planet with wavelength. The layout of the paper is as follows. In Section 2 we present the observations. Section 3 discusses our data reduction methods. In Section 4 we report the analysis of the data, measure the physical properties of the system, construct the planet's transmission spectrum, and compare this theoretical predictions. We conclude in Section\,5.
 
%==============================================================================================================
\section{Observations}
%==============================================================================================================

We monitored the flux of {\rm{WASP$-$23}} (also known as GSC 07635-01376, $\alpha= 06^{{\rm h}} 44^{{\rm m}} 30.65^{{\rm s}}$, $\delta=-42^{{\circ}} 45^{{\prime}} 41.0^{{\prime\prime}}$; J2000.0; $V=12.68$ mag) during one transit on February 25, 2012 (Fig.~1 \& 2). The data were collected using the GROND instrument, attached to the MPG/ESO 2.2m telescope at ESO La Silla Observatory (Chile)\footnote{Detailed information on the 2.2m telescope, the GROND instrument and the relevant technical settings can be found at http://www.eso.org/sci/facilities/lasilla/telescopes/2p2/}. GROND (\textbf{G}amma \textbf{R}ay Burst \textbf{O}ptical and \textbf{N}ear-Infrared \textbf{D}etector) is a simultaneous multi-channel imager that has been specifically designed for gamma-ray burst (GRB) afterglow observations \citep{g8}. Its capability to simultaneously monitor sources in seven optical to NIR passbands make it an excellent instrument for obtaining high quality, multi-band transit light curves. It is capable of yielding light curves with a root-mean-square (r.m.s.) scatter better than $1$\,mmag, depending on the atmospheric conditions \citep{lendl2010, nikolov2012, mancini2013}.

%%%%%%%%%%%%%%%%%%%%%%%%%%%%%%%%%%%%%%%%%%
%                                                FIGURE 2                                                       %
%%%%%%%%%%%%%%%%%%%%%%%%%%%%%%%%%%%%%%%%%%
   \begin{figure*}[!t]
   \centering
   \includegraphics[width=\hsize]{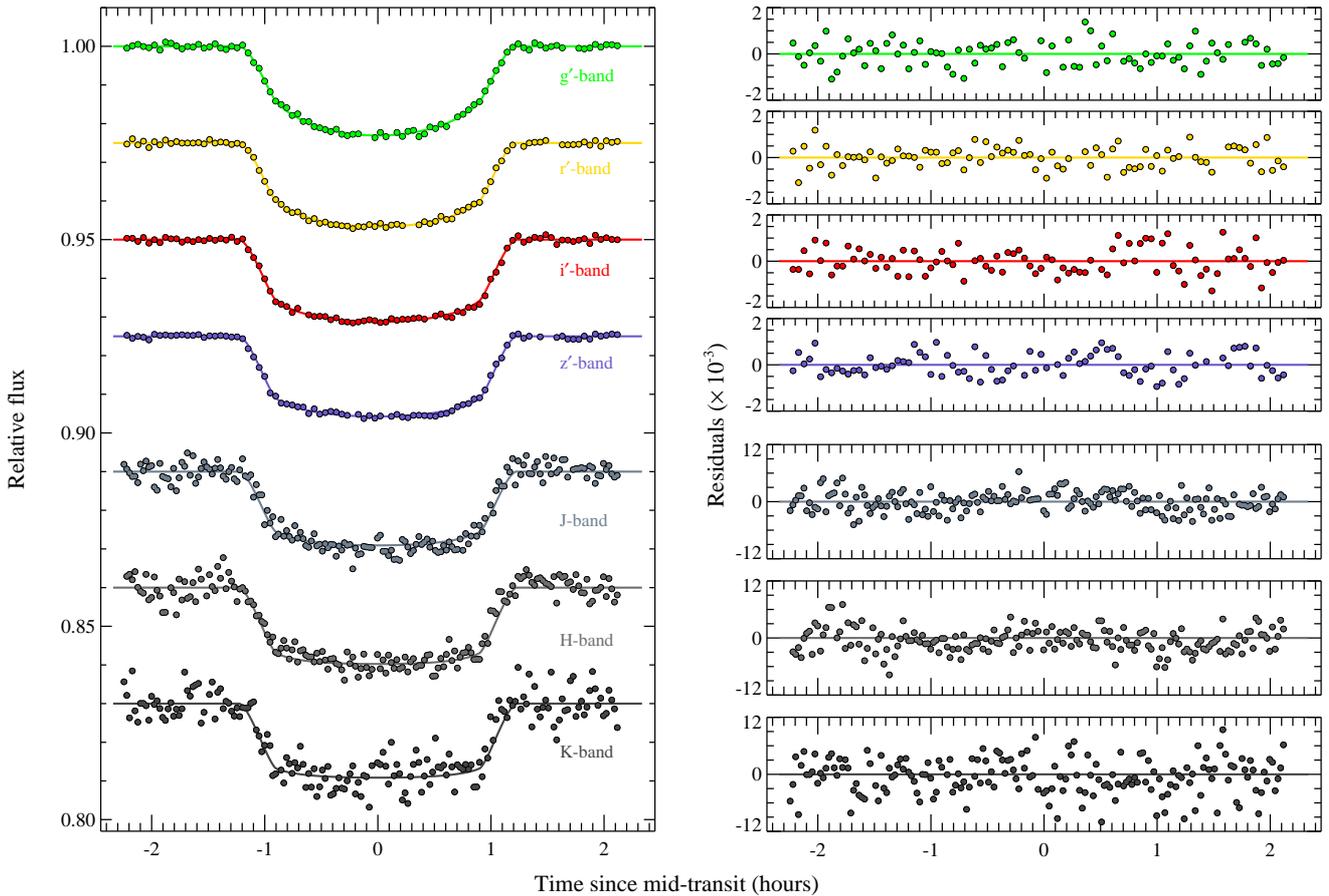}
      \caption{{\it{Left panel:}} Detrended transit light curves (dots) of and the best-fit models (lines) for the {{WASP$-$23}} GROND data. The optical photometry clearly exhibits a change in the shape of the transit due to the different limb-darkening in each band. {\it{Right panel:}} Residuals of the light curve compared to the fitted models.}
         \label{FigVibStab}
   \end{figure*}
%%%%%%%%%%%%%%%%%%%%%%%%%%%%%%%%%%%%%%%%%%%

To reach GROND's detectors the light from the telescope is split using dichroics into seven beams, resulting in response curves close to the Sloan g$^\prime$ ($\lambda = $ 459 nm), r$^\prime$ ($\lambda =$ 622 nm), i$^\prime$ ($\lambda =$ 764 nm), z$^\prime$ ($\lambda =$ 899 nm) in the optical and J ($\lambda =$ 1256 nm), H ($\lambda =$ 1647 nm) and K ($\lambda =$ 2151 nm) bands in the NIR, respectively. The optical beams are incident onto four $2048 \times 2048$ back-illuminated E2V CCDs with $5.4\arcmin\times5.4\arcmin$ fields of view (FOVs) at a plate scale of $0\arcsec.158$ pixel$^{-1}$. The NIR channels are imaged onto three $1024 \times 1024$ Rockwell HAWAII-1 arrays with $10\times10$ arcmin FOVs and plate scales 0$\arcsec$.6 pixel$^{-1}$. The off-axis guide camera was used to autoguide the telescope. 

Light curves were acquired in GROND's non-dither mode to decrease systematics associated with inter-pixel sensitivity variations. We obtained data for {\rm{WASP$-$23}} and eleven and seven reference stars in the optical and NIR images, respectively. We applied a moderate defocus to the 2.2m telescope, such that the point spread functions (PSFs) had diameters $\sim76$ pixels ($12\arcsec$) in the optical and $\sim13$ pixels ($7.\arcsec6$) in NIR. Table 1 gives details of the instrumental setup and illumination levels of the seven beams. As the guide camera was also defocussed, we made sure in advance that at least one bright star was within its FOV. 
%%%%%%%%%%%%%%%%%%%%%%%%%%%%%%%%%%%%%%%%%%
%                                                  TABLE 1                                                       %
%%%%%%%%%%%%%%%%%%%%%%%%%%%%%%%%%%%%%%%%%%
\begin{table}
\caption{Instrumental and observational settings used to obtain 
data for {\rm{WASP$-$23}} (median values), during the 
observation on UT February 25, 2012.}             % title of Table
\label{table:1}      % is used to refer this table in the text
\centering                          % used for centering table
\begin{tabular}{c c c c c}        % centered columns (4 columns)
\hline\hline                 % inserts double horizontal lines
GROND & Gain & $d{\tablefootmark{a}}_{{\rm ring}}$  & Max count & Background \tablefootmark{b}  \\    % table heading 
band       &   (ADU)  &  (pixel)           &  (ADU)$\times 10^{3}$    &   (ADU)$\times 10^{3}$  \\    % table heading 
\hline                        % inserts single horizontal line
   g'      &   1.45   &     57  &  18.6  &    0.3 \\      % inserting body of the table
   r'        &    1.33 &     67  &  43.8  &   0.3 \\
   i'         &   1.62 &     51  &  27.2  &    0.3 \\
   z'       &    1.74 &     48  &  22.3  &   0.4 \\
   J         &   2.39 &       9  &  26.7  &  5.0 \\ 
   H        &   2.65 &       6  &  18.2  & 17.0 \\
   K         &   2.55 &     24  &  13.1  &  11.0 \\ 
   \hline                                   %inserts single line
\end{tabular}
\tablefoot{\tablefoottext{a}{Diameter of the defocussed psf ring in pixels.}
\tablefoottext{b}{Measured using an annulus, centered on the stellar psf, encompassing an area four times larger than the area used to produce a photometric measurement  of the stellar psf.}}
\end{table}
%%%%%%%%%%%%%%%%%%%%%%%%%%%%%%%%%%%%%%%%%%

The main advantages of the telescope defocus technique have been discussed by several researchers, e.g. \cite{winn2007}, \cite{gillon_2007}, \cite{s10} (and references therein), \cite{c10}. Although the method results in non-uniformly defocused PSFs of the stars across large FOVs ($\geq 1^\circ \times 1^\circ$), the technique performs well on small FOVs (such as GROND). It allows superb PSF sampling, significantly reducing the random and systematic errors resulting from detector imperfections and resposne variations. The technique also avoids detector saturation due to fluctuations in seeing, because such fluctuations are small compared to the level of defocusing. We note that telescope defocusing is not well suited to photometry in crowded fields or when the target star has nearby companions. We used the 2MASS catalogue to rule out these possibilities for our observations of WASP-23.

%%%%%%%%%%%%%%%%%%%%%%%%%%%%%%%%%%%%%%%%%%
%                                                  TABLE 2                                                       %
%%%%%%%%%%%%%%%%%%%%%%%%%%%%%%%%%%%%%%%%%%
\begin{table}
\caption{r'-band light curve extract}             % title of Table
\label{table:1}      % is used to refer this table in the text
\centering                          % used for centering table
\begin{tabular}{c c c c c c}        % centered columns (4 columns)
\hline\hline                 % inserts double horizontal lines
BJD\tablefootmark{a} & Flux & $\sigma_{{\rm Flux}}$\tablefootmark{b}  & $z$\tablefootmark{c} & $x$\tablefootmark{d}  & $y$\tablefootmark{e}\\    % table heading 
\hline                        % inserts single horizontal line
  2.52787  &  1.000410  &  0.000256   & 1.03539  &   7.73  &  71.17 \\    
  2.52990  &  1.001160  &  0.000269   & 1.03370  &   7.61  &  70.30 \\     
  2.53193  &  0.999446  &  0.000253   & 1.03274  &   7.54  &  70.06  \\    
  2.53396  &  1.000615  &  0.000255   & 1.03170  &   6.95  &  70.66  \\    
  2.53599  &  0.998938  &  0.000253   & 1.03100  &   6.72  &  70.67  \\    
  2.53802  &  1.000043  &  0.000252   & 1.03079  &   6.18  &  69.93  \\    
   \hline                                   %inserts single line
\end{tabular}
\tablefoot{\tablefoottext{a}{BJD -- 2455980;}
\tablefoottext{b}{1-$\sigma$ error-bar of the flux, as measured by APER;}
\tablefoottext{c}{airmass;}
\tablefoottext{d}{x -- 630 horizontal pixel position of the defocussed psf;}
\tablefoottext{d}{y -- 1300 vertical pixel position of the defocussed psf;}}
\end{table}
%%%%%%%%%%%%%%%%%%%%%%%%%%%%%%%%%%%%%%%%%%

We observed {\rm{WASP$-$23}} on February 25 2012 for 4.4\,hr from UT 00h32m to 04h56m. An exposure time of 90\,s was used in the optical and resulted in 91 integrations. The slow read-out mode was selected ($\sim48$\,s read-out-time) as it delivers high-quality images \footnote{The GROND detector can also be read-out in `fast' mode ($\sim11$\,s read-out-time), but it has been found during previous transit observations that images obtained in that mode suffer from systematic effects that lower the quality of the resulting light curves.}. In the NIR averaged stacks of seven 10\,s integrations were obtained to improve the signal-to-noise ratio (SNR), resulting in a total of 181 images. At the beginning of the observation the airmass of the target began at 1.04, peaked at 1.03, and increased to 1.47 by the end of the observing sequence. The observing conditions were nearly photometric, with seeing variation from $\sim16$ to $\sim18$ pixels as measured from a fit to the wings of the defocused PSFs.

Theoretical models of exoplanet transmission spectra imply that the NIR region is richer than the optical in terms of potential spectroscopic signatures. We therefore aimed to obtain high-precision light curves for that region as well. The control of GROND allows only a single integration time (limited from 0 to 10\,s) for the three JHK bands simultaneously, so we optimised the exposure time such that the highest SNR was obtained for the band when the star had the most flux (J). For the degree of telescope defocus, we aimed to optimise the SNR in the optical, as we expected better performance in that region based on previous experience \citep{nikolov2012, mancini2013}.

%==============================================================================================================
\section{Data reduction and analysis}
%==============================================================================================================

The data were reduced and analysed on a channel-by-channel basis, using a customised pipeline. GROND's optical and NIR data are provided in separate fits files. The images from the optical channels are packed into four-extension data cubes and the JHK read-outs are delivered as a single $3072\times1024$ image. After processing to collapse the optical data cubes and to split each NIR image, the data were sorted into science and calibration images.  

For the optical data a median-combined master bias frame was computed using 20 zero-second images. We estimated the average dark current of GROND's optical detectors to be $\sim0.01\rm{e}^{-} s^{-1}\rm{pixel}^{-1}$, which is negligible for our data. We therefore neglected the dark current correction. A median-combined master flat-field was calculated using six bias-corrected and dithered twilight sky flat-fields that survived a selection criterion for linearity in each of the passbands. Finally, each optical science frame was de-biased and flat-field corrected to produce calibrated time series of images.

In the NIR, a series of ten dark frames with integration time of 10\,s were employed to compute a median-combined master dark frame. A series of 24 dithered twilight sky flat frames were used to obtain a median-combined master sky flat. The master dark frame was subtracted from all the science and sky flat frames. Each dark-subtracted science image was then corrected with the master sky flat frame. An electronic odd-even readout pattern along the y-axis was removed before flat-fielding. This was done by smoothing each image and comparing it with the unsmoothed one after the master dark had been subtracted. The amplitudes of the read-out pattern were then determined by comparing the median level of each column to the overall median level. Each column was then shifted to the overall median level. The master sky flats were divided out from the science images after removal of the read-out pattern.

Previous investigators have commonly obtained dithered images prior to the main observing sequence to further correct the science images against sky background variations. However, similar corrections are inadequate for sky gradients that vary on timescales less than the duration of the observations. In addition, relative photometry, if performed at gradient-stable sky backgrounds is in practice insensitive to sky variations that concern simply the sky level. Finally, previous analyses of {\rm{WASP$-$44\,b}} \citep{mancini2013} with and without sky corrections resulted in indistinguishable light curves with similar scatter. We therefore did not obtain dithered sky images. 

%%%%%%%%%%%%%%%%%%%%%%%%%%%%%%%%%%%%%%%%%%
%                                                  TABLE 3                                                       %
%%%%%%%%%%%%%%%%%%%%%%%%%%%%%%%%%%%%%%%%%%
\begin{table}
\caption{Theoretical limb-darkening coefficients, 
light curve residual scatter (in parts per million, ppm) and the $\beta-$factor
used to estimate the photometric uncertainties of the 
GROND data.}             % title of Table
\label{table:1}      % is used to refer this table in the text
\centering                          % used for centering table
\begin{tabular}{c c c c c}        % centered columns (4 columns)
\hline\hline                 % inserts double horizontal lines
GROND   & $u_1$ &  $u_2$ &  r.m.s   &  $\beta$\\    % table heading 
band        &            &             &  (ppm)  &             \\    % table heading 
\hline                        % inserts single horizontal line
g$^\prime$  &  0.7413  &   0.0719    &    419  & 1.00   \\ 
r$^\prime$  &  0.5323  &   0.1804     &    371  & 1.00\\
i$^\prime$     &  0.4224  &   0.1983  &    430   & 1.27\\
z$^\prime$   &  0.3491  &   0.2062   &    330   & 1.28\\
J                   &  0.1063  &   0.2589   &  1480   & 1.50\\
H                  &  0.1126  &   0.3130   & 1760   &   1.52\\
K                  &  0.2389  &   0.2467   & 2980   &   1.16\\
   \hline                                   %inserts single line
\end{tabular}
\end{table}
%%%%%%%%%%%%%%%%%%%%%%%%%%%%%%%%%%%%%%%%%%

We performed circular aperture photometry on each of the calibrated images. To accurately determine the central positions of the heavily defocused PSFs, we proceeded as follows. First, we extracted sub-images, roughly centred on each star. The brightest stars were rings with typical diameters $\sim50$ and $\sim15$ pixels in the optical and NIR, respectively. We then convolved each image with a Gaussian kernel to produce a smoothed PSF with a well-defined peak, located at the centre of the original PSF. Finally, we fitted a two-dimensional Gaussian to measure the PSF centre and full width at half maximum, which was latter used to monitor the seeing variations during the observation. Instrumental fluxes and uncertainties (computed taking into account a random photon noise) were acquired using a {\sc daophot}-type photometry algorithm as implemented in the {\sc idl}\footnote{The acronym IDL stands for Interactive Data Language. For further details see http://www.exelisvis.com/ ProductsServices/IDL.aspx} {\sc aper}\footnote{part of {\sc aper} Astronomy User's Library. For detailed information see http://idlastro.gsfc.nasa.gov/} routine.  

To compute the relative flux light curve of {\rm{WASP$-$23}} in the four optical channels we used 2MASS\,J06444546-4242083 as a comparison star. Because of the relatively small FOV of GROND we performed an exhaustive investigation of the possible reference stars prior to the observations. This object satisfied the requirements, being isolated enough to avoid contamination with bright neighbours when defocused, and being the only stellar source in the optical FOV of a similar brightness and colour to {\rm{WASP$-$23}}, thus reducing systematic effects due to differential colour extinction. For instance, the instrumental magnitude differences ({\rm{WASP$-$23}} minus comparison star), measured on UT Februray 25 2012 were $\Delta g^{\prime}=0.510$\,mag, $\Delta r^{\prime}=0.082$\,mag,  $\Delta i^{\prime}=0.319$\,mag,  $\Delta z^{\prime}=0.785$\,mag. For the same reasons, in the NIR channels we used 2MASS J06445925-4245061 as the comparison star. The instrumental magnitude differences were $\Delta J=-0.784$\,mag, $\Delta H=-0.696$\,mag and $\Delta K=-0.396$\,mag. 

To select optimum sizes for the inner aperture and sky annulus we investigated the root-mean-square (rms) of the out-of transit baseline flux as a function of various combinations of aperture and annulus sizes. Table 2 exhibits an r'-band light curve extract along with the airmass and measurements of the $(x,y)$ drift of the PSF on the detector employed to correct the light curve during the analysis.

We used the UT time stamps available in the fits headers of the data and computed the central times of each integration using the exposure times. All times were then converted to Barycentric Julian Date (BJD), using the converter of Eastman et al.\ (2010).

%%%%%%%%%%%%%%%%%%%%%%%%%%%%%%%%%%%%%%%%%%
%                                                 FIGURE 2                                                       %
%%%%%%%%%%%%%%%%%%%%%%%%%%%%%%%%%%%%%%%%%%
   \begin{figure}[!h]
   \centering
   \includegraphics[width=\hsize]{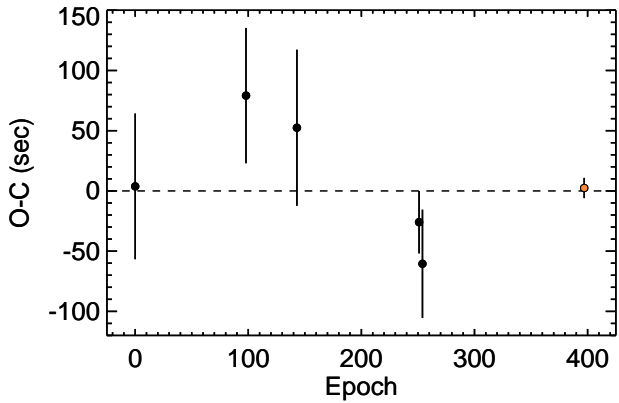}
      \caption{Observed minus computed (O-C) transit times 
      for \object{{\rm{WASP$-$23\,b}}}, using observations reported in the literature
      (black dots) and our measurement (orange dot),
      compared to a linear ephemeris (dashed line).}
         \label{FigVibStab}
   \end{figure}
%%%%%%%%%%%%%%%%%%%%%%%%%%%%%%%%%%%%%%%%%%

%==============================================================================================================
\section{Light curve analysis} %\label{bozomath}
%==============================================================================================================
We performed the light curve analysis in two steps. First, we simultaneously fitted the seven optical to NIR GROND light curves to derive accurate estimates of the transit central time $T_{c}$ and orbital period $P$, and evaluated the orbital parameters and planetary radius ($i$, $a/R_{\ast}$, $R_{p}/R_{\ast}$).
 
In the second step we complemented the GROND photometry with the radial velocity (RV) measurements of {\rm{WASP$-$23}}, reported by \cite{t11} and performed a joint photometry plus RV MCMC analysis using {\sc exofast} \citep{eastman2012}. As our ultimate goal is to construct a ground-based transmission spectrum of WASP-23, based on uniform transit parameters from all bands we excluded the RVs obtained during transit, which are affected by the Rossiter-McLaughlin effect.

%%%%%%%%%%%%%%%%%%%%%%%%%%%%%%%%%%%%%%%%%%
%                                                  TABLE 4                                                       %
%%%%%%%%%%%%%%%%%%%%%%%%%%%%%%%%%%%%%%%%%%
\newcommand{\bjdtdb}{\ensuremath{\rm {BJD_{TDB}}}}
\newcommand{\feh}{\ensuremath{\left[{\rm Fe}/{\rm H}\right]}}
\newcommand{\teff}{\ensuremath{T_{\rm eff}}}
\newcommand{\ecosw}{\ensuremath{e\cos{\omega_*}}}
\newcommand{\esinw}{\ensuremath{e\sin{\omega_*}}}
\newcommand{\msun}{\ensuremath{\,{\rm M}_\odot}}
\newcommand{\rsun}{\ensuremath{\,{\rm R}_\odot}}
\newcommand{\lsun}{\ensuremath{\,{\rm L}_\odot}}
\newcommand{\mj}{\ensuremath{\,{\rm M}_{\rm J}}}
\newcommand{\rj}{\ensuremath{\,{\rm R}_{\rm J}}}
\newcommand{\fave}{\langle F \rangle}
\newcommand{\fluxcgs}{10$^9$ erg s$^{-1}$ cm$^{-2}$}

\begin{table}
\caption{Derived system parameters for \object{{\rm{WASP$-$23\,b}}} from light curve fits, compared to 
the corresponding results reported in the discovery paper. The final estimates of our analysis are 
the weighted mean values from all bands.}
\label{table:1}
\centering
\begin{tabular}{l c c c}
\hline\hline
Band & $R_{p}/R_{\ast}$ & $i^{\circ}$ & $a/R_{\ast}$ \\
\hline
GROND $g^{\prime}$  &  $0.13402_{-0.00061}^{+0.00062}$  &  $87.93_{-0.22}^{+0.24}$  &  $9.900_{-0.14}^{+0.14}$\\
GROND $r^{\prime}$  &  $0.13336_{-0.00044}^{+0.00045}$  &  $87.85_{-0.21}^{+0.22}$  &  $9.850_{-0.13}^{+0.14}$\\
GROND $i^{\prime}$  &  $0.13433_{-0.00043}^{+0.00044}$  &  $88.02_{-0.26}^{+0.30}$  &  $9.910_{-0.16}^{+0.17}$\\
GROND $z^{\prime}$  &  $0.13400_{-0.00041}^{+0.00042}$  &  $87.83_{-0.24}^{+0.25}$  &  $9.850_{-0.15}^{+0.15}$\\
GROND $J$  &  $0.1340_{-0.0011}^{+0.0012}$  &  $88.12_{-0.74}^{+0.93}$  &  $10.24_{-0.49}^{+0.44}$\\
GROND $H$  &  $0.13520_{-0.0012}^{+0.0012}$  &  $87.75_{-0.70}^{+0.91}$  &  $9.970_{-0.52}^{+0.56}$\\
GROND $K$  &  $0.1326_{-0.0014}^{+0.0014}$  &  $88.74_{-1.0}^{+0.85}$  &  $10.59_{-0.63}^{+0.32}$\\
\hline
Weighted mean &    $0.13394_{-0.00022}^{+0.00022}$  &  $87.91_{-0.11}^{+0.12}$  &  $9.893_{-0.070}^{+0.071}$\\
\hline
Triaud et al. (2011) &  $0.13004_{-0.00045}^{+0.00040}$ & $88.39_{-0.45}^{+0.79}$  & $10.61$\\
\hline
\end{tabular}
\end{table}
%%%%%%%%%%%%%%%%%%%%%%%%%%%%%%%%%%%%%%%%%%

As {\sc exofast} allows an analysis of one light curve at a time we use the results from the first step (the simultaneous fit to the photometry) as priors for {\sc exofast}. The latter was used to derive reliable estimates of the physical properties and their uncertainties for \object{{\rm{WASP$-$23\,b}}} and its parent star. {\sc exofast} utilizes empirical polynomial relations \citep{torres_et_al2010} between the masses and radii of stars, and their surface gravities ($\log g$), effective temperatures ($T_{\rm{eff}}$) and metallicities [Fe/H], based on a large sample of well-studied non-interacting binaries. Including this empirical relation as a constraint allows a derivation of the full orbital and physical properties from transit light curves and RV data \citep{eastman2012}.

In the first step, we modelled the transit light curves simultaneously in flux units). A two-component {\bf{correction}} function was constructed, containing a transit model multiplied by a baseline function that accounts for the most significant known systematic effects affecting ground-based transit observations: airmass and detector intra-pixel sensitivity variations (Fig. 1). 

The first component is based on the analytical transit models of \cite{Mandel_Agol}, which in addition to the transit central time ($T_{c}$) and orbital period ($P$) is a function of the orbital inclination ($i$), planet semi-major axis and radius normalized to the stellar radius ($a/R_{\ast}$) and ($R_{p}/R_{\ast}$), respectively. To account for the limb darkening of the star we adopted the quadratic limb darkening law:

\begin{equation}
\frac{I_{\mu}}{I_{1}} = 1 - u_{1}(1-\mu) - u_{2}(1 - \mu)^{2},
\end{equation}
 
where $I$ is the intensity and $\mu$ is the cosine of the angle between the line of sight and the normal to the stellar surface. We fixed the two limb darkening coefficients ($u_{1}$ and $u_{2}$) to their theoretical values (see Table 3 for details), which we obtained by interpolating within the calculated and tabulated ATLAS models \citep{Clar_boem, eastman2012} to the stellar parameters $T_{\mathrm{eff}} = 5150\pm100$ K, $\log \, g  = 4.4\pm0.2$ (cgs), $[\mathrm{Fe/H}] = -0.05\pm0.13$ and $v_{\mathrm{t}} = 0.8\pm0.3$ km $\rm s^{-1}$, which we adopted from \cite{t11}.

We constructed a baseline function to account for systematic effects that concern the (in- and out-of-transit) photometry. It contains a linear airmass ($z$) term as well as a quadratic polynomial of the PSF location $(x,y)$ on the detector to take into account for any inter-pixel sensitivity variations:

\begin{equation}
F(z,x,y) = 1+a_{0}+a_{1}z+a_{2}x+a_{3}x^{2}+a_{4}y+a_{5}y^{2}+a_{6}xy.
\end{equation}

To derive the best-fit parameters we employed the Levenberg-Marquardt least-squares algorithm, minimising the $\chi^2$ function over the data (Fig. 2) \citep{Marq2009}. Conveniently, {\sc exofast} also takes into account the systematics, incorporating a baseline function and trends, as described in \cite{eastman2012}. We therefore included function (3) in {\sc exofast} to take into account the systematics and to enable a uniform analysis of the in- and out- of-transit photometry.\footnote{We choose to employ {\sc mpfit} in the first step as it allows one to construct a joint model incorporating the entire dataset. In contrast {\sc exofast} permits an analysis on a per-channel basis, i.e.\ one band at a time. We therefore, determined universal values for $a/R_{\ast}$, $i$, $T_c$, $P$ and $R_{p}/R_{\ast}$ using the complete dataset and used these results as priors in {\sc exofast}, where the fitting is done for each band.} 

An accurate derivation of system parameters from transit light curve fits requires a realistic estimation of the related photometric errors. The photometric uncertainties produced by {\sc idl}/{\sc aper} are usually underestimates, so we ignored them in our analysis. These are formal statistical uncertainties and do not take into account time-correlated noise. We used the out-of-transit data of each light curve before detrending and applied the `time-averaging' procedure proposed by \citet{pont2006} and used by various authors including \cite{gillon2006}, \cite{winn2007, winn2008}, \cite{gibson2008}, \cite{nikolov2012}, \cite{s12a, s12b} and \cite{mancini2012} to inflate the photometric errors to more realistic values. In summary, for each band we computed the $\beta-$ratio between the standard deviation of the out-of-transit unbinned and binned data with bin sizes similar to the transit ingress/egress duration. The photometric uncertainties were therefore set to the out-of-transit unbinned standard deviation multiplied by the corresponding $\beta-$factor (see Table 3).

%==============================================================================================================
\subsection{System parameters} %\label{bozomath}
%==============================================================================================================

To derive system parameters for {\rm{WASP$-$23}}, including the priors for {\sc exofast}, we relied on the light curves from the seven GROND passbands. We assumed that $i$ and $a$ should not depend on the observed band and constrained them to a single universal value. In contrast, we treated the planetary radius as a free parameter for each passband. We also set the linear and the quadratic limb darkening coefficients to their theoretical values as well as the initial guess for the orbital period to the value reported in \cite{t11}. We found $i= 88.02^{\circ}\pm0.19^{\circ}$, $a/R_{\ast} = 9.94\pm0.11$ and ratios of the radii for each passband:
$R_{p}/R_{\ast}(g^{\prime})=0.13410\pm 0.00066$,
$R_{p}/R_{\ast}(r^{\prime})=0.13348\pm 0.00049$,
$R_{p}/R_{\ast}(i^{\prime})=0.13455\pm 0.00047$,
$R_{p}/R_{\ast}(z^{\prime})=0.13439\pm 0.00048$, 
$R_{p}/R_{\ast}(J)=0.1340\pm  0.0016$, 
$R_{p}/R_{\ast}(H)=0.1352\pm  0.0016$, 
$R_{p}/R_{\ast}(K)=0.1325\pm  0.0020$.
These values were input in {\sc exofast} as priors along with the star's $T_{\mathrm{eff}}$, $\log{g}$ and [Fe/H], fitting the transit photometry and RVs simultaneously in each band. The results for the main physical properties are summarised in Table 4 \& 5. Our results are slightly different from these reported in \cite{t11}, though with higher precision, favouring a more inflated planet. In particular we improved the precision of $i$ by factors of $\sim$4 and $\sim$6.5 for the lower and upper limits, respectively. We find the radius of {\rm{WASP$-$23b}} to be $R_{p}=1.067_{-0.038}^{+0.045} R_{Jup}$, which is larger than the radius presented in the discovery paper. 
 
%%%%%%%%%%%%%%%%%%%%%%%%%%%%%%%%%%%%%%%%%%
%                                                  TABLE 5                                                       %
%%%%%%%%%%%%%%%%%%%%%%%%%%%%%%%%%%%%%%%%%%
\begin{table*}
\caption{Stellar and planetary parameters for {\rm{WASP$-$23}} derived from light curve and RV fits to the GROND photometry and RV measurements from \cite{t11}.}
\label{table:1}
\centering
\begin{tabular}{l c c c}
\hline\hline
Quantity & Symbol & This work & \cite{t11}\\
\hline
\multicolumn{3}{l}{\it{Stellar parameters}} \\
\hline
Mass (\msun) & $M_{*}$  & $0.842_{-0.049}^{+0.051}$   &  $0.78_{-0.12}^{+0.13}$\\
Radius (\rsun) & $R_{*}$ & $0.819_{-0.028}^{+0.033}$  &  $0.765_{-0.049}^{+0.033}$ \\
Luminosity (\lsun) & $L_{*}$ & $0.421_{-0.046}^{+0.053}$  & -\\
Density (cgs) & $\rho_*$ & $2.112_{-0.044}^{+0.046}$    & $1.843_{-0.027}^{+0.025}$ \\
Surface gravity (cgs) & $\log(g_*)$  & $4.531_{-0.029}^{+0.025}$ & $4.4\pm{0.2}$ \\
Effective temperature (K) & $\teff$ & $5149_{-93}^{+93}$  & $5150\pm{100}$ \\
Metallicity  & $\feh$  & $-0.06_{-0.13}^{+0.13}$  & $-0.05\pm0.13$  \\
\hline
\multicolumn{3}{l}{\it{Planetary parameters}} \\
\hline
Semi-major axis (AU) & $a$  & $0.03798_{-0.00075}^{+0.00075}$  & $0.0376_{-0.0024}^{+0.0016}$\\
Mass (\mj) & $M_{P}$ & $0.917_{-0.039}^{+0.040}$   &  $0.884_{-0.099}^{+0.088}$\\
Radius (\rj)  & $R_{P}$ & $1.067_{-0.038}^{+0.045}$  & $0.962_{-0.056}^{+0.047}$ \\
Density (cgs) & $\rho_{P}$ & $0.916_{-0.102}^{+0.096}$  &  -\\
Surface gravity & $\log(g_{P})$  & $3.294_{-0.031}^{+0.027}$  & - \\
Equilibrium temperature$\tablefootmark{a}$ (K) &$T_{eq}$  & $1152_{-26}^{+28}$  & - \\
Safronov Number  & $\Theta$ & $0.0770_{-0.0035}^{+0.0033}$  &  -\\
Incident flux (\fluxcgs) & $\fave$   & $0.400_{-0.035}^{+0.040}$ & - \\
Orbital period (day)& $P$   & $2.9444300\pm0.0000011$ & $2.9444256^{+0.0000011}_{-0.0000013}$ \\
Reference transit time (BJD) & $T_{0}$   & $2454813.68159\pm0.00042$ & $2455320.12363^{+0.00012}_{-0.00013}$ \\

\hline
\multicolumn{3}{l}{\it{RV parameters}} \\
\hline
RV semi-amplitude (m/s) & $K$  & $145.57_{-2.49}^{+2.44}$    & - \\
%Minimum mass (\mj) & $M_P\sin i$ & $0.916_{-0.039}^{+0.040}$  & - \\
Mass ratio   & $M_{P}/M_{*}$  & $0.001039_{-0.000027}^{+0.000027}$  & - \\
\hline
\hline
\end{tabular}
\tablefoot{\tablefoottext{a}{Assuming a little redistribution over the surface of the planet, i.e., the area that is reemitting with average 
temperature $T_{eq}$ is $2\pi R_{p}^2$, as in \cite{hansen2007}.}}
\end{table*}%%%%%%%%%%%%%%%%%%%%%%%%%%%%%%%%%%%%%%%%%%

%==============================================================================================================
\subsection{Ephemeris} %\label{bozomath}
%==============================================================================================================

One of our main goals was to secure the transit ephemeris ($T_c$ and $P$) of \object{{\rm{WASP$-$23\,b}}}. We fixed all physical parameters to their values derived in the previous section and fitted GROND's light curves simultaneously to derive a transit time. We added this to all reported central transit times in the literature (Table 6), then performed a linear fit as a function of the observed epoch ($E$):

\begin{equation}
T_{\mathrm{C}}(E) = T_{\mathrm{o}} + E\times P.
\end{equation}       

We derived $P=2.9444300\pm0.0000011$\,d and $T_c = 2454813.68159\pm0.00042$ BJD. This period is in poor agreement ($\sim$$3\sigma$ level) with the value reported in \cite{t11}  ($P = 2.9444256_{-0.0000013}^{+0.0000011}$\,d), but is more reliable due to the larger time span of transit times we used. We investigated the observed minus calculated (O-C) residuals of the timings versus our ephemeris. We found no systematic deviation and conclude that there is no compelling evidence for transit timing variations (Fig. 3).

%%%%%%%%%%%%%%%%%%%%%%%%%%%%%%%%%%%%%%%%%%
%                                                  TABLE 6                                                       %
%%%%%%%%%%%%%%%%%%%%%%%%%%%%%%%%%%%%%%%%%%
\begin{table}
\centering
\caption{Transit mid-times and O-C residuals computed using transit 
light curves from the literature and the observation reported in 
this work. Note the high precision ($\sim$8s) of the the transit time derived using GROND's light curves.}
\begin{tabular}{@{} r c c c }
\hline
Epoch & Central time & O$-$C & Refernce\\
 & (${\rm{BJD}}_{{\rm{TDB}}}$) & (days) & \\
\hline
    0 & $245   4813.68163^{ 0.00070}_{ 0.00073}$ &   0.000044 &        1\\
   98 & $245   5102.23664^{ 0.00065}_{ 0.00068}$ &   0.000916 &        1\\
  143 & $245   5234.73568^{ 0.00075}_{ 0.00079}$ &   0.000607 &        1\\
  251 & $245   5552.73321^{ 0.00030}_{ 0.00030}$ &  -0.000300 &        1\\
  254 & $245   5561.56610^{ 0.00052}_{ 0.00051}$ &  -0.000700 &        1\\
  397 & $245   5982.620315^{ 0.000097}_{ 0.000097}$ &   0.000028 &        2\\
\hline
\end{tabular}
\tablefoot{$1-$ \cite{t11}; $2-$ this work.}
\end{table}
%%%%%%%%%%%%%%%%%%%%%%%%%%%%%%%%%%%%%%%%%%

%%%%%%%%%%%%%%%%%%%%%%%%%%%%%%%%%%%%%%%%%%
%                                                 FIGURE 3                                                       %
%%%%%%%%%%%%%%%%%%%%%%%%%%%%%%%%%%%%%%%%%%
   \begin{figure}[!b]
   \centering
   \includegraphics[width=\hsize]{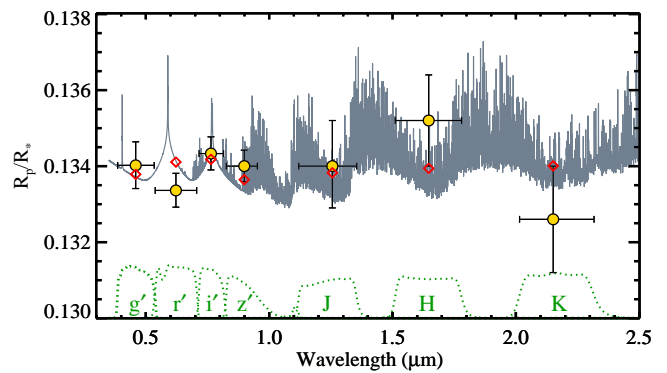}
      \caption{Radius variation (measured with $R_{p}/R_{\ast}$ and indicated with 
      yellow disks with error bars) of \object{{\rm{WASP$-$23\,b}}} in the seven optical to NIR
      passbands, compared to a synthetic transmission spectrum based on a planet-wide 
      pressure-temperature profile (grey continuous line). The open red diamonds
      indicate the expected theoretical values, integrated over GROND's filter curves. The latter
      are displayed as dotted lines at the origin of the graph.}
         \label{FigVibStab}
   \end{figure}
 %%%%%%%%%%%%%%%%%%%%%%%%%%%%%%%%%%%%%%%%%%

%%%%%%%%%%%%%%%%%%%%%%%%%%%%%%%%%%%%%%%%%%
%                                                 FIGURE 4                                                       %
%%%%%%%%%%%%%%%%%%%%%%%%%%%%%%%%%%%%%%%%%%
   \begin{figure}[!b]
   \centering
   \includegraphics[width=\hsize]{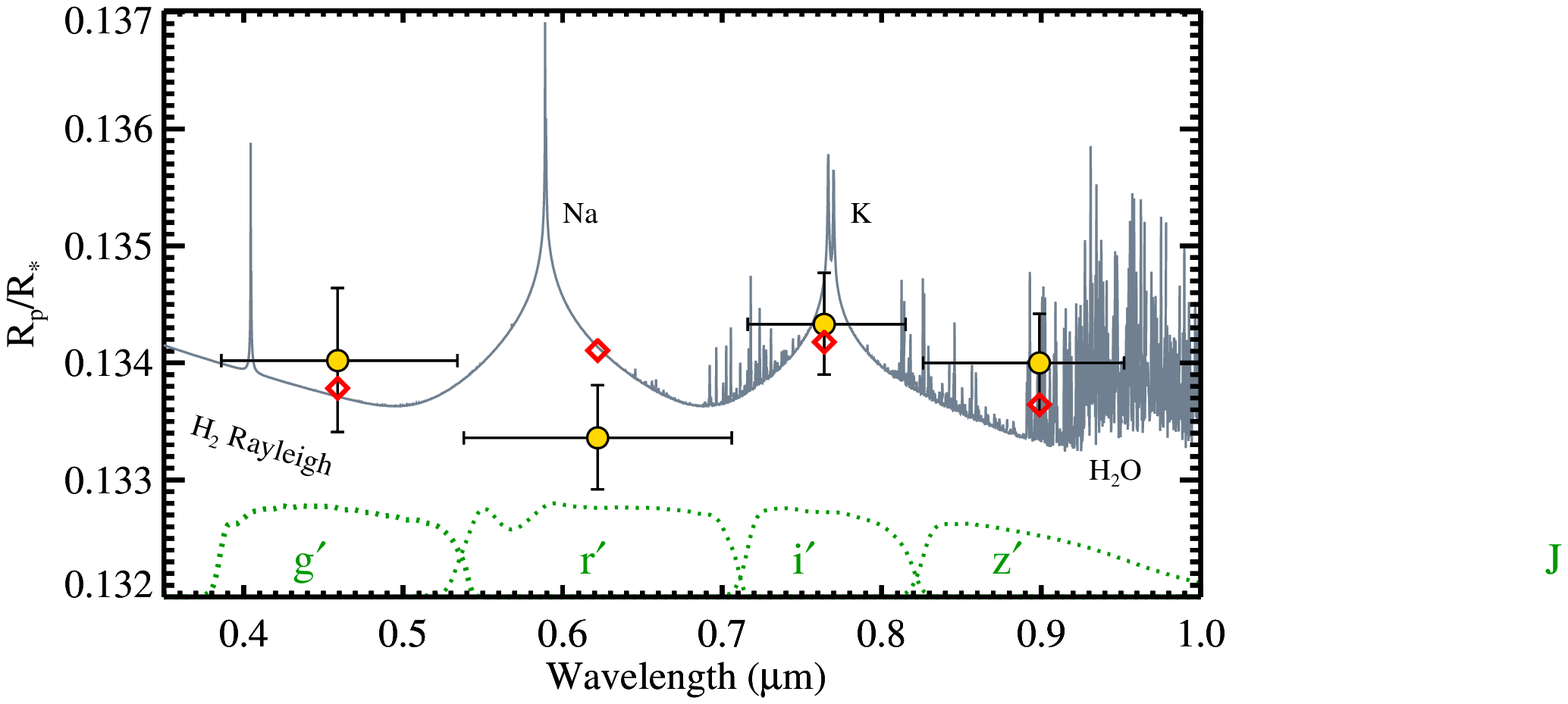}
      \caption{Same as Figure 4, but displaying a closer view of the radius variation 
      in the optical bands.}
         \label{FigVibStab}
         \label{FigVibStab}
   \end{figure}
%%%%%%%%%%%%%%%%%%%%%%%%%%%%%%%%%%%%%%%%%%

%==============================================================================================================
\subsection{Radius Variation} 
%==============================================================================================================

A set of simultaneous multi-band transit light curves provides an opportunity to investigate the behaviour of the measured planetary radius with wavelength. Atmospheric absorption features in the optical regime due to sodium (Na) and potassium (K) have been theoretically predicted \citep{seager_sasselov2000} and observed in the transmission spectra of hot Jupiters, e.g.\ XO-2b \citep{sing2011b, sing2012} from the ground. Contrary to this example of an agreement between theory and observation, an analysis of {\it{HST}} ultraviolet-optical spectrophotometric data of HD\,189733\,b by \cite{sing2011a} revealed a featureless spectrum, which might be an indication for the presence of high altitude clouds. These observational results indicate the diversity of exoplanet atmospheres, the understanding of which requires more observational results.  

To investigate the radius variation of \object{{\rm{WASP$-$23\,b}}} in GROND's passbands we fixed $P$, $i$ and $a/R_{\ast}$ to their final results from the analysis in Section 4.1, and the linear and quadratic limb darkening coefficients to their predicted values. We fitted for the planetary radius (specifically the ratio of the radii $R_{\rm{p}}/R_{\rm{\ast}}$) and the instrumental systematics. This approach improves the precision of the measured planetary radius, by avoiding sources of uncertainty common to all measurements of $R_{\rm{p}}/R_{\rm{\ast}}$. In addition, the fit takes into account the information for the system parameters and reflects differences in the transit depth that are determined only by the size of the planet in each passband. To estimate realistic values for the normalised planet radii and their error bars we carried out the fitting using an MCMC approach.   The results for the planet radii, obtained at this step are displayed in Figs.\ 4 and 5. The vertical error bars comprise the 68.3\% confidence levels from the MCMC distributions of $R_{p}/R_{\ast}$. As expected, the radius measurements in the optical regime show much smaller error bars compared to the NIR passbands. This is due to the lower scatter of the data in the optical. We also show horizontal bandwidths (plotted with error-like bars) that have been computed as the full widths at half maximum, using GROND's filter curves, which are indicated at the lower part of the plot. 

To compare our observational results with theorical predictions we employed the system parameters from Table 5 and computed a 1D model atmosphere of \object{{\rm{WASP$-$23\,b}}}, using the code described in \cite{fortney2005, fortney2008}. The fully non-gray model uses the chemical equilibrium abundances of \cite{lodders2002} and the opacity database described in \cite{freedman2008}. The atmospheric pressure-temperature profile simulates planet-wide average conditions. We computed the transmission spectrum of the model using the methods described in \cite{fortney2010}. For each of the GROND passbands we computed the expected theoretical value given the synthetic transmission spectrum, integrating over the corresponding filter curve. Finally, we used the derived theoretical values for the planet radius and fitted them as a model to the measured values of $R_{p}/R_{\ast}$ from the seven GROND light curves (Fig. 4 \& 5). We find a good agreement between the observationally derived measurements and synthetic values of the planetary radii, consistent with a featureless transmission spectrum at the $\sim$1$\sigma$ level. Two of our radius estimates ($r^{\prime}$-band in the optical and $H$-band in the NIR) were found to be off by slightly more than 1$\sigma$ from their corresponding theoretical values. This is most likely due to systematics of unknown origin. 

%==============================================================================================================
\section{Summary and discussion} %\label{bozomath}
%==============================================================================================================

We have presented observational results based on data collected with the GROND instrument on the MPG/ESO 2.2m telescope at ESO La Silla, during one transit of the hot Jupiter \object{{\rm{WASP$-$23\,b}}}, the first photometric follow-up of this object after its discovery. We employed the telescope defocusing technique and autoguiding to optimise the quality of the light curve and minimise the instrumental systematics in the complete optical to NIR capability of the GROND instrument for the first time. Light curves constructed from this data set showed a scatter versus fitted models as low as 330 ppm (Table 3). For comparison, a recent observation performed by \cite{tregloran} with the 3.58\,m New Technology Telescope (NTT), in defocussed mode obtained light curve scatters of 258 and 211 ppm, during two transit events of {\rm{WASP$-$50\,b}}. Their lower scatter is expected due to the larger aperture of the telescope.

We used the full potential of GROND's optical to NIR dataset to refine the ephemeris and system parameters of the {\rm{WASP$-$23b}} system and compared our results with these presented by \cite{t11}. We complemented our photometric data with existing RV measurements of the parent star to derive the full physical properties of the system. We measured the central transit time with a precision $\sim$8\,s, complemented this measurement with published transit epochs, and refined the orbital period of \object{{\rm{WASP$-$23\,b}}}. We analysed the O-C diagram of the transit times and found no significant evidence for transit timing variations.  Using all of the seven GROND bands we determined a larger and more precise value for the radius of \object{{\rm{WASP$-$23\,b}}}, which is in agreement at the $\sim$2$\sigma$ level with that from \cite{t11}. The latter however, was derived based on a joint fit to five lower-quality light curves covering the $z$, $R$ and {\it{I}}$+${\it{z}} passbands. Despite our larger value for the planetary radius, \object{{\rm{WASP$-$23\,b}}} still sits below the region of the so-called heavily ``{\em bloated}'' hot Jupiters, in accord to the theoretical models for irradiated giant planets of \cite{baraffe2008}. Moreover, \cite{fortney_et_al_2007a} presented theoretical mass-radius relationships for irradiated giant exoplanets, parametrized by mass, age, and an effective orbital distance, defined as the distance from the Sun, where a hypothetical planet would receive the same flux as the actual planet. Given the luminosity of {\rm{WASP$-$23}} and the semi-major axis of its planet we evaluated that distance to be $a=0.585$\,AU. We then interpolated the models of \cite{fortney_et_al_2007a} to match the distance and the mass of {\rm{WASP$-$23b}} for a solar composition at 4.5\,Gyr and evaluated $R_{p}=1.042$ ${\rm{R}}_{Jup}$, which is within the error of our result.
  
Finally, we used the potential of the GROND light curves to investigate the (possible) variation of the planetary radius with wavelength and constructed a ground-based simultaneous optical to NIR broad-band transmission spectrum of \object{{\rm{WASP$-$23\,b}}}. According to our results for the system parameters (Table 5) the equilibrium temperature of \object{{\rm{WASP$-$23\,b}}} is $T_{eq}=1152_{-26}^{+28}$\,K, which places the planet among the pL-class (cooler) of giant close-in exoplanet atmospheres in the scheme proposed by \cite{fortney2008}. Taking into account these results, we generated a synthetic atmospheric transmission model of an irradiated hot Jupiter, assuming a planet-wide pressure-temperature profile and compared theory with observations. The optical part of the synthetic transmission spectrum at this equilibrium temperature and surface gravity (20.73\,${\rm{m\ s^{-2}}}$) suggests a radius variation due to the presence of gaseous sodium and potassium in the atmosphere of \object{{\rm{WASP$-$23\,b}}}, while the NIR parts are dominated by water vapour features. The detection of such radius variation should be a realistic task, if observations could be taken at the precision levels attained in this work and with a higher spectral resolution. However, even the narrowest passband of GROND ($i^{\prime}$-band) spans $\sim$120\,nm, which effectively reduces the sensitivity of our measurements to radius variations. This effect is illustrated in Figs.\ 4 and 5, where using the filter curves of GROND we computed the theoretically expected values for the radius of \object{{\rm{WASP$-$23\,b}}}, given the synthetic spectrum. We fitted our observational result for the radius to the theoretical estimates. Although we found the $r^{\prime}$-band, which covers the expected sodium absorption line, to deviate from the predicted theoretical value, overall the theoretical model is in very good agreement with observational transmission spectrum at the $\sim$1$\sigma$ level. 

%==============================================================================================================
\begin{acknowledgements}
The authors acknowledge Timo Anguita and R\'{e}gis Lachaume for their
technical support at the time of the observations, David Sing and Fr\'ed\'eric Pont
for their constructive discussions as well as the anonymous referee for the detailed and constructive
suggestions that improved the manuscript.
N.N. acknowledges support from an STFC consolidated grant.  
J.S. acknowledges STFC for the award of an Advanced Fellowship.
\end{acknowledgements}
%==============================================================================================================

%==============================================================================================================

\end{document}